\documentclass[pre,aps,groupaddress,showpacs,prd,twocolumn]{revtex4}

\usepackage{epsfig,amsmath,graphicx,amssymb}
\def\be{\begin{equation}}
\def\ee{\end{equation}}
\def\bee{\begin{eqnarray}}
\def\ene{\end{eqnarray}}
\def\bes{\begin{subequations}}
\def\ees{\end{subequations}}

\newcommand{\tw}{ \tilde{w}}

\begin{document}

\title{Thresholdless surface solitons}

\author{Yuliy V. Bludov$^1$, Yaroslav V.  Kartashov$^2$, and Vladimir V. Konotop$^3$}

\address{ $^1$ Centro de F\'{\i}sica, Universidade do Minho,
Campus de Gualtar, Braga 4710-057, Portugal
\\
$^2$ICFO-Institut de Ciencies Fotoniques, and Universitat Politecnica de Catalunya, Mediterranean Technology Park, 08860 Castelldefels (Barcelona), Spain
\\
 $^3$Centro de F\'{\i}sica Te\'orica e Computacional and Departamento de F\'{\i}sica, Faculdade de Ci\^encias, Universidade de Lisboa,
Avenida Professor Gama
Pinto 2, Lisboa 1649-003, Portugal
}


\begin{abstract}

We report on the existence of nonlinear surface waves which, on the one hand, do not require the threshold energy flow for their excitation, and, on the other hand, extend into media at both sides of the interface at low powers, i.e. can not be reduced to the conventional Tamm states. Such waves can be excited  if the refractive index in at least one of the materials forming the interface is periodically modulated, with properly selected modulation depth and frequency. Thresholdless surface solitons can be stable in the entire existence domain.

\end{abstract}

\maketitle

\noindent

The history of the surface waves goes back to the seminal paper of Tamm~\cite{Tamm}, who upon consideration of the Kronig-Penney model for an electron in a crystal, has shown that at the interface between a periodic and  homogeneous structures  there can exist localized states, which cannot be supported by a surface between two homogeneous media.  Four decades later, in Ref.~\cite{Yeh}  Tamm's ideas were introduced in the optics of periodic structures, where localized states of the electric field at the boundary between a homogeneous and stratified media were predicted.
 At the same time it was understood~\cite{Tomlinson}  that the nonlinearity can dramatically change the situation leading to new surface states, whose analogs do not exist in the linear structures. Nowadays the issue of the nonlinear surface modes attracts growing  attention (see e.g. the recent review~\cite{review} and the references there in).
In particular, it was suggested that nonlinear surface modes may form at the interface of periodic and uniform nonlinear media \cite{ra}. Such states were thoroughly studied and observed in both focusing \cite{rb}, and defocusing~\cite{rc,rd,re} media.

One of the main features of the nonlinear surface states at the boundary between two  semi-infinite media, studied so far, stems form the fact that  they have no linear limit:  such modes possess the energy flow threshold below which they do not exist.  Such thresholds have been widely discussed in the literature, see e.g.~\cite{ra, rb,rc,rd,re}.
So far, thresholdless surface states were observed only at the interface of two periodic structures separated by a suitable interval~\cite{LinTam1,SurfSolit}, or at the interfaces of lattices with defect surface channel~\cite{rf,rg}, {or at the interfaces of non-periodic lattices~\cite{c:bessel}}. Notice however, that in all these settings surface modes bifurcate from well-localized linear defect states (i.e. from the Tamm states).

In the present Letter we show that there exists one more type of the nonlinear surface modes, which on the one hand do not require threshold energy flow for their excitation and on the other hand do not have linear localized counterparts, {unlike all surface modes reported so far}.

We consider  beam propagation  which can be described by the equation for the dimensionless electric field $q$:
\begin{eqnarray}
iq_\xi = -q_{\eta\eta} + R(\eta)q+\sigma|q|^2q.
\label{GPEq}
\end{eqnarray}
Here  $\xi$ and $\eta$ are the  longitudinal and  transverse  coordinates normalized to the diffraction length and the characteristic beam width. It will be assumed that the plane $\eta=0$ separates two media. The right hand side medium ($\eta>0$) is considered periodic with the modulation of the refractive index described by $R_{right}(\eta)= R_{r}\cos(2\eta)$ (we emphasize that the particular choice of the cos-like  refractive index  is for the illustration purposes only, and the results reported below are valid
for any other type of refractive index modulation). For the refractive index of media on the left hand side, $\eta<0$, we suppose $R_{left}(\eta)= R_{l,0}+R_{l,1}\cos(2\eta)$ and consider the two different examples: of the homogeneous Kerr medium ($R_{l,1}=0$) and periodically modulated one, $R_{l,1}\ne 0$. We concentrate on the cases of two focusing, $\sigma=-1$, or defocusing, $\sigma=1$ media (the generalization of our arguments to the case of the interface between focusing and defocusing media is straightforward).

\begin{figure}
  \begin{center}
   \begin{tabular}{c}
       \hspace{-2.5cm}
       \includegraphics[width=1.1\columnwidth]{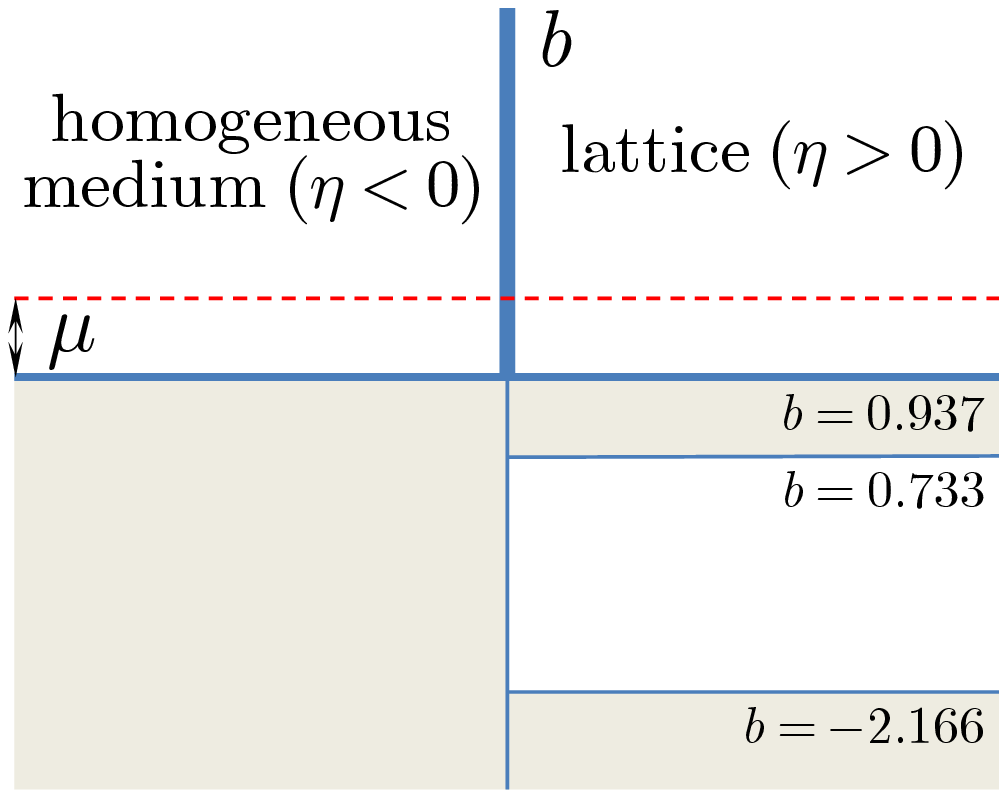}
       \hspace{-5.5cm}
       \includegraphics[width=1.1\columnwidth]{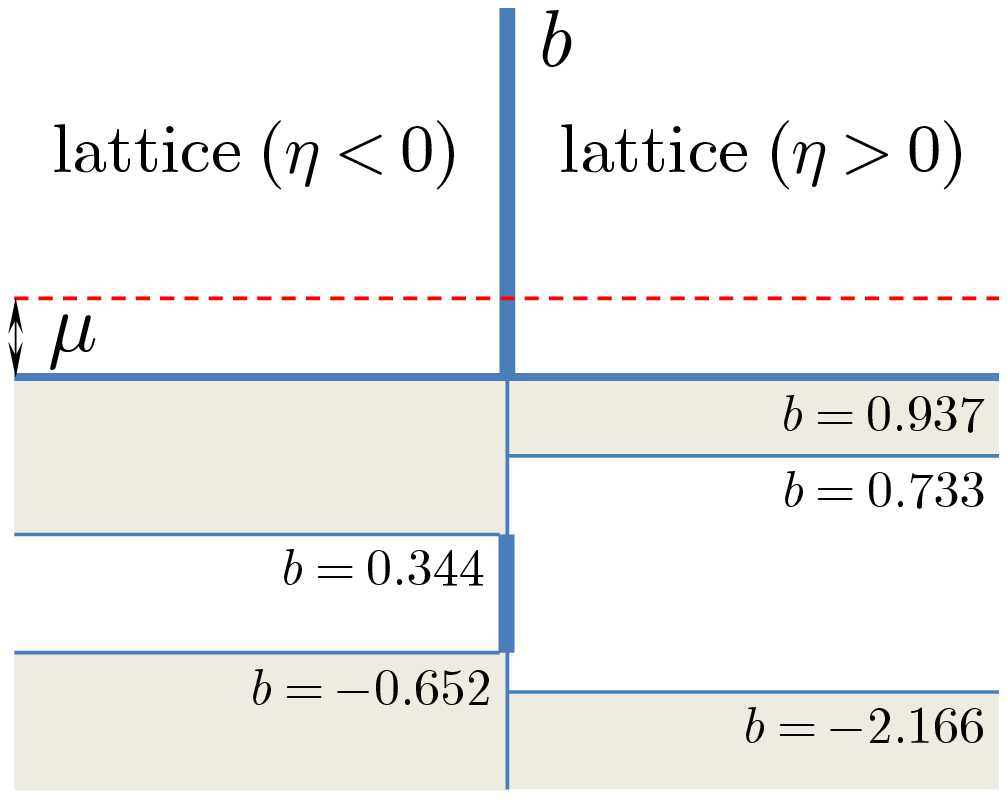}
   \end{tabular}
   \end{center}
\vspace{-3cm}
\caption{(Color online) Schematic representation of interfaces between homogeneous and periodic media (left panel) and between two different periodic media (right panel). Shadowed areas show the linear spectrum (allowed bands). The bold parts of the axis $b$ indicate the intervals to which the propagation constant must belong for existence of any surface soliton. The bold horizontal line shows the edge of the linear spectrum, from which the thresholdless solitons bifurcate.   The dashed lines show examples of location of the propagation constants of the surface solitons.  {The values of $b$ indicated  in the left and right panels correspond to the specific examples shown below in Figs.~\ref{fig:ne} and \ref{fig:ne-tm}}.
}
\label{fig:1}
\end{figure}

To explain the main idea we start with the case, where $R_{left}(\eta)= R_{l,0}$. We are interested in stationary solutions $q=e^{ib\xi}w(\eta)$ (here $b$ is the real propagation constant and $w(\eta)$ describes the soliton profile) bifurcating from the linear spectrum.
In order to get insight into properties of such states it is instructive to  start from the linear limit $q\to 0$ (or formally $\sigma=0$). To this end we concentrate on the linear equation $-b\tw(\eta) = -\tw_{\eta\eta} + R(\eta)\tw(\eta)$, obtained from (\ref{GPEq}) by neglecting the nonlinear term, which at $\eta\to\infty$ decays more rapidly than the linear ones (notice that $w(\eta)\to\tw(\eta)$ at $\eta\to \infty$). Now we recall the standard arguments giving the Tamm mode (see e.g.~\cite{Anselm}). The linear solution for $\eta<0$ is given by $\tw_l(\eta)=C_l\exp\left(\mu_l\eta\right)$ where $\mu_l=\sqrt{R_{l,0}+b}$, while the solution at $\eta>0$ has the form of the Bloch function $\tw_r=P(\eta)\exp\left(-\mu_r \eta\right)$ where $P(\eta)$ is a $\pi$-- or $2\pi$--periodic function and $\mu_r$ is the  Floquet exponent (we have taken into account that the propagation constant $b$ belongs to one of the   gaps of the spectrum of the lattice $R_{right}(\eta)$). Then the condition for existence of the Tamm mode in the general case reads $\mu_r=P_\eta(0)/P(0)-\mu_l$.
In our particular case, where  $\max_\eta R_{right}(\eta)=R_{right}(0)$, we have $P_\eta(0)=0$ and thus $\mu_r\to 0$, at $b\to -R_{l,0}$. In this limit we also have that $\tw_{r,\eta}(0)\to 0$ (hereafter the subscript $\eta$ stands for the derivative, i.g. $w_{l,\eta}=dw_l/d\eta$). Since, in a periodic medium, the Floquet exponent tends to zero only if the propagation constant approaches one of the gap edges, we conclude that the limit $\mu_r\to 0$ is possible  only if  $R_{right}(\eta)$ is such that $-R_{l,0}$ coincides with one of the boundaries of the gaps (or allowed zones).   This leads us to the first condition, necessary for existence of the thresholdless surface solitons:  the boundary of the  $b$-domain   where  the localized states exist in the homogeneous medium must coincide with one of the boundaries of the stop gaps of the periodic medium. This situation is illustrated in Fig.~\ref{fig:1}(a). (If $P_\eta(0)\neq 0$, i.e. the interface between the two structures does not coincide with one of the extrema of the periodically modulated refractive index, the respective corrections must be introduced in the consideration.)

Turning now to the nonlinear case, we consider the limit $b\to -R_{l,0}$ (illustrated by the dashed lines in Fig.~\ref{fig:1})
when surface solitons are very wide and their envelopes inside right (periodic) medium can be rather accurately described by the homogeneous nonlinear Schr\"odinger (NLS) equation for the envelope of the Bloch wave, with the effective diffraction $D=-d^2b_n(k)/dk^2|_{k=0,1}$ where $b_n(k)$ is the $n$-th band of dispersion relation of  $R_{right}(\eta)$ ($k$ is the Bloch wavevector; $k=0$ and $k=1$ correspond to the center and the boundary of the Brillouin zone).

First we consider the case of the defocusing media ($\sigma=1$).
Then soliton can exist only if $D<0$ and its maximum can be located only in the right (periodic) medium. For $\eta<0$ the solution is given by $w_l=\sqrt{2}\mu_l/\sinh\left(\mu_l(\eta_l-\eta)\right)$, where $\eta_l>0$ is a   constant, which must be found from the continuity of $w$ and $w_\eta$ at the surface. When  $\eta=0$, the function $w_l(\eta)$ (defined at $\eta\leq 0$) achieves its maximum $w_{l,max}=\sqrt{2}\mu_l/\sinh\left(\mu_l \eta_l\right)$. In this point one verifies $w_{l,\eta}\sim \mu_l^2/\sinh^2\left(\mu_l \eta_l\right)\sim w_l^2$ for $\mu_r\to 0$. On the other hand, for $\eta\geq 0$, where $D>0$, the envelope of the Bloch state is described by the $\cosh^{-1}$-like function, and at $\eta=0$ one has the relation $w_{r, \eta}(0)\sim w_r(0)$ for $\mu_l\sim\mu_r\to 0$. Since, there must also verify $w_l(0)=w_r(0)$ and  $w_{l,\eta}(0)=w_{r,\eta}(0)$, the above asymptotics for $w_{l,r}$ are not compatible. This means that the amplitude of the mode  does not go zero,
i.e. no thresholdless surface solitons can exist.

Consider now the case of the focusing media ($\sigma=-1$). Now for $\eta<0$ the solution is given by $w_l=\sqrt{2}\mu_l/\cosh\left(\mu_l(\eta-\eta_l)\right)$. For $\eta\geq 0$ one must require now $D<0$ (notice that, if $D>0$, the solution in the periodic media is of $\sinh^{-1}$-like type, and the above arguments about existence of thresholdless solitons at the interface between focusing and effectively defocusing media, inhibit the existence of thresholdless modes). Now the asymptotics of the left-medium and right-medium solutions are compatible, i.e. the limit $b\to -R_{l,0}$ can allow for existence of the solution localized about the surface.   Thus we conclude, that the solution, localized in the vicinity of surface, can exist  when $D>0$, what is satisfied for all upper (lower) boundaries of the allowed (forbidden) zones of the lattice. With this assertion we now can formulate the requirement for the existence of the thresholdless surface solitons in a focusing medium: {\it the boundary of the existence domain of the localized states in the homogeneous medium must coincide with one of the upper boundaries of the allowed bands}.

\begin{figure}
  \begin{center}
   \begin{tabular}{c}
       \includegraphics[width=\columnwidth]{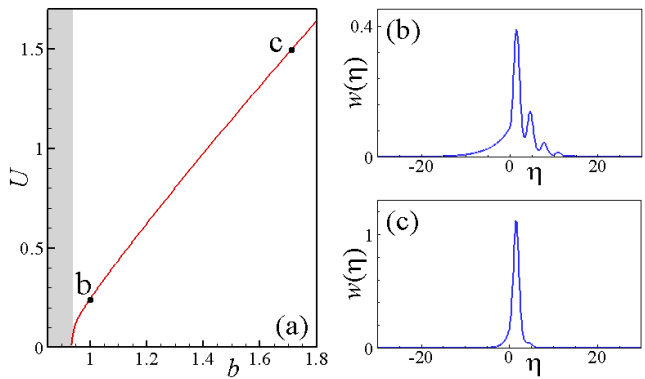}
   \end{tabular}
   \end{center}
\vspace{0.3cm}
\caption{(Color online) (a) Energy flow \textit{vs} propagation constant  for the interface between periodic $R_r=3$ and homogeneous $R_{l,0}=-0.9368$ media for the focusing ($\sigma=-1$) nonlinearity. Shaded area corresponds to the first allowed band of the lattice spectrum; (b,c) shape of surface soliton at corresponding points of panel (a) for  $b=0.94$ (b) and  $b=1.7132$ (c). The zone structure for situation is schematically depicted in the left panel of Fig.~\ref{fig:1}.}
\label{fig:ne}
\end{figure}

\begin{figure}
  \begin{center}
   \begin{tabular}{c}
       \includegraphics[width=\columnwidth]{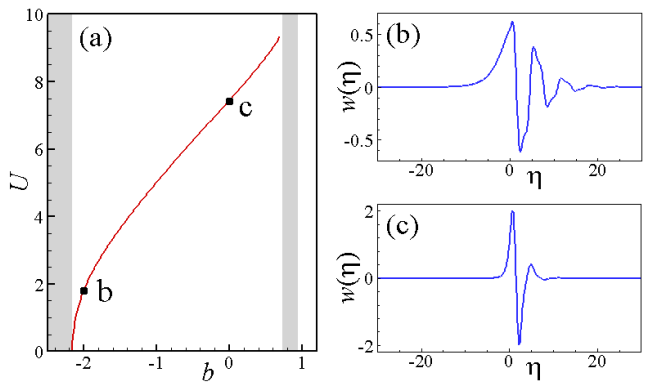}
   \end{tabular}
   \end{center}
\vspace{0.3cm}
\caption{(Color online) (a) Energy flow \textit{vs} propagation constant for the interface between periodic $R_r=3$ and homogeneous $R_{l,0}=2.166$ media for the focusing nonlinearity. Left and right shaded areas correspond to the first   and second allowed bands; (b) and (c) shapes of surface solitons at the corresponding points of panel (a) for  $b=-2.15$  and $b=0$. }
\label{fig:ne-1gap}
\end{figure}

The numerical analysis of this case is illustrated in  Fig.~\ref{fig:ne}(a), where the branch of the solution is depicted in the plane $(b,U)$ where $U=\int_{-\infty}^{\infty}|w|^2d\eta$ is the total energy flow along the surface.  We consider the case corresponding to Fig. 1(a) when the lower boundary of existence domain inside the uniform medium $b=-R_{l,0}$ coincides with lower boundary of semi-infinite gap in right medium. One observes that the soliton branch indeed bifurcates from the continuous spectrum, i.e.   surface solitons can be excited at any infinitely small power. The surface soliton width decreases with the increase of soliton energy flow, as it follows from the comparison of Figs.~\ref{fig:ne}(b) and \ref{fig:ne}(c). Such solitons are completely stable {(we observe that for them $dU/db>0$,  similar to the stability of solitons in a homogeneous medium ensured by the  Vakhitov-Kolokolov criterion)}. The same situation occurs when propagation constant falls into first gap of periodic right medium, when the $-R_{l,0}$ coincides with the upper edge of the second allowed band (see Fig.\ref{fig:ne-1gap}).  These solutions however are unstable.

\begin{figure}
  \begin{center}
   \begin{tabular}{c}
       \includegraphics{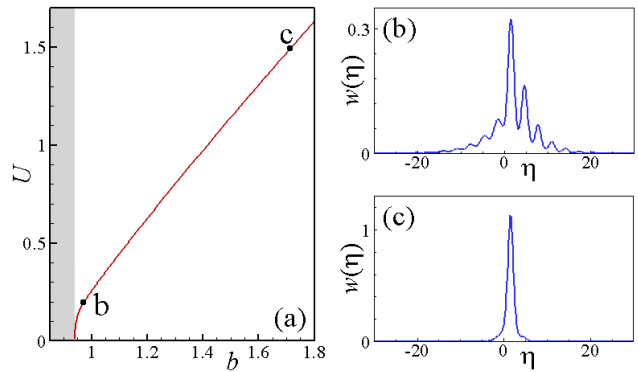}
   \end{tabular}
   \end{center}
\caption{(Color online) (a) Energy flow \textit{vs} propagation constant  for the interface between periodic media with parameters $R_r=3$, $R_{l,0}=-0.815$, $R_{l,1}=1$ media for the focusing media. Shaded area corresponds to the first band of the spectrum; (b,c) shapes of surface solitons at corresponding points of panel (a) for  $b=0.9456$ (b)  and  $b=1.7183$ (c).The zone structure for this situation is schematically depicted in the right panel of Fig.~\ref{fig:1}. }
\label{fig:ne-tm}
\end{figure}

Let us now turn to a boundary between two dissimilar lattices, which we studied for the focusing case $\sigma=-1$. Employing the above arguments  one can conclude that for existence of thresholdless surface solitons the coincidence of two upper boundaries of the allowed bands of the left and right lattices is required. A representative example of soliton family whose propagation constant emerges from semi-infinite gaps (with coinciding lower edges) of both lattices is shown in Fig.~\ref{fig:ne-tm}.
Such solitons are stable too in the entire domain of their existence.

To conclude, we have reported a possibility of existence of the thresholdless surface solitons, which in the linear (i.e. small energy flow) limit are transformed
into small-amplitude extended states approaching Bloch waves inside the periodically modulated medium and having long slowly decaying tails inside the uniform medium.
In this sense the reported surface modes differ drastically from conventional surface waves possessing excitation thresholds and from modes reducing in the linear limit to localized Tamm states.

The main condition for existence of the thresholdless solitons is matching of the boundaries of the linear spectra at the both sides of the interface. In practical terms the required conditions can be implemented in various ways,
e.g., by adjusting of geometrical properties of the lattices (periods, depths of modulations, etc.), by using temperature gradients (i.e. change of the refractive index by  heating one of the two media), by doping of the materials that results in modifications in refractive index, etc.


\begin{thebibliography}{99}

\bibitem{Tamm} I. E. Tamm, Z. Phys. {\bf 76} 849 (1932).

\bibitem{Yeh} P. Yeh, A. Yariv, and A. Y. Cho, Appl. Phys. Lett. {\bf 32}, 2776 (2006)
104 (1978)

\bibitem{Tomlinson} W. J. Tomlinson,  Opt. Lett. {\bf 5}, 323 (1980).

\bibitem{review} Y. S. Kivshar,   Laser Phys. Lett. {\bf 5}, 703 (2008).

\bibitem{ra} K. G. Makris, S. Suntsov, D. N. Christodoulides, G. I. Stegeman, and A. Hache,   Opt. Lett. {\bf 30}, 2466 (2005).

\bibitem{rb}	S. Suntsov, K. G. Makris, D. N. Christodoulides, G. I. Stegeman, A. Hach\'e, R. Morandotti, H. Yang, G. Salamo, and M. Sorel,  Phys. Rev. Lett. {\bf 96}, 063901 (2006).

\bibitem{rc}  Y. V. Kartashov, V. A. Vysloukh, and L. Torner, Phys. Rev. Lett. {\bf 96}, 073901 (2006).

\bibitem{rd} E. Smirnov, M. Stepic, C. E. Rüter, D. Kip, and V. Shandarov, Opt. Lett. {\bf 31}, 2338 (2006).

\bibitem{re} C. R. Rosberg, D. N. Neshev, W. Krolikowski, A. Mitchell, R. A. Vicencio, M. I. Molina, and Y. S. Kivshar,  Phys. Rev. Lett. {\bf 97}, 083901 (2006).

\bibitem{LinTam1} S. Suntsov, K. G. Makris, D. N. Christodoulides, G. I. Stegeman, R. Morandotti,
M. Volatier, V. Aimez, R. Ar\'es, C.  E. R\"uter, and D. Kip,  Opt. Express {\bf 16}, 4663 (2007)

\bibitem{SurfSolit} S. Suntsov, K. G. Makris, D. N. Christodoulides, G. I. Stegeman, R. Morandotti,
M. Volatier, V. Aimez, R. Ar\'es, E. H. Yang,  and G. Salamo,    Opt. Express {\bf 16}, 10480 (2008)


\bibitem{rf} A. Szameit, Y. V. Kartashov, M. Heinrich, F. Dreisow, T. Pertsch, S. Nolte, A. T\"unnermann, F. Lederer, V. A. Vysloukh, and L. Torner,   Opt. Lett. {\bf 34}, 797 (2009).

\bibitem{rg} N. Malkova, I. Hromada, X. Wang, B. Garnett, and Z. Chen,  Opt. Lett. {\bf 34}, 1633 (2009).

\bibitem{c:bessel} X. Yang, L. Dong, and G. Yin,   Appl. Phys. B {\bf 95}, 179 (2009).

\bibitem{Anselm} A. I. Anselm, "Introduction to Semiconductor Theory" (Prentice Hall, 1982)



\end{thebibliography}
\end{document}